# Revealing in-plane grain boundary composition features through machine learning from atom probe tomography data

Xuyang Zhou[1,2*], Ye Wei[1], Markus Kühbach[1,3], Huan Zhao[1], Florian Vogel[4], Reza Darvishi Kamachali[5], Gregory B. Thompson[2], Dierk Raabe[1], Baptiste Gault[1,6]

[1] Max-Planck-Institut für Eisenforschung GmbH, 40237 Düsseldorf, Germany.

[2] Department of Metallurgical & Materials Engineering, The University of Alabama, 35401 Tuscaloosa, USA.

[3] Fritz-Haber-Institut der Max-Planck-Gesellschaft, 14195 Berlin, Germany.

[4] Institute of Advanced Wear & Corrosion Resistant and Functional Materials, Jinan University, 510632 Guangzhou, China.

[5] Federal Institute for Materials Research and Testing (BAM), 12205 Berlin, Germany.

[6] Department of Materials, Royal School of Mines, Imperial College London, SW7 2AZ London, UK.

*Correspondence to: x.zhou@mpie.de.

**Abstract:** Grain boundaries (GBs) are planar lattice defects that govern the properties of many types of polycrystalline materials. Hence, their structures have been investigated in great detail. However, much less is known about their chemical features, owing to the experimental difficulties to probe these features at the atomic length scale inside bulk material specimens. Atom probe tomography (APT) is a tool capable of accomplishing this task, with an ability to quantify chemical characteristics at near-atomic scale. Using APT data sets, we present here a machine-learning-based approach for the automated quantification of chemical features of GBs. We trained a convolutional neural network (CNN) using twenty thousand synthesized images of grain interiors, GBs, or triple junctions. Such a trained CNN automatically detects the locations of GBs from APT data. Those GBs are then subjected to compositional mapping and analysis, including revealing their in-plane chemical decoration patterns. We applied this approach to experimentally obtained APT data sets pertaining to three case studies, namely, Ni-P, Pt-Au, and Al-Zn-Mg-Cu alloys. In the first case, we extracted GB specific segregation features as a function of misorientation and coincidence site lattice character. Secondly, we revealed interfacial excesses and in-plane chemical features that could not have been found by standard compositional analyses. Lastly, we tracked the temporal evolution of chemical decoration from early-stage solute GB segregation in the dilute limit to interfacial phase separation, characterized by the evolution of complex composition patterns. This machine-learning-based approach provides quantitative, unbiased, and automated access to GB chemical analyses, serving as an enabling tool for new discoveries related to interface thermodynamics, kinetics, and the associated chemistry-structure-property relations.

**Graphical Abstract:**

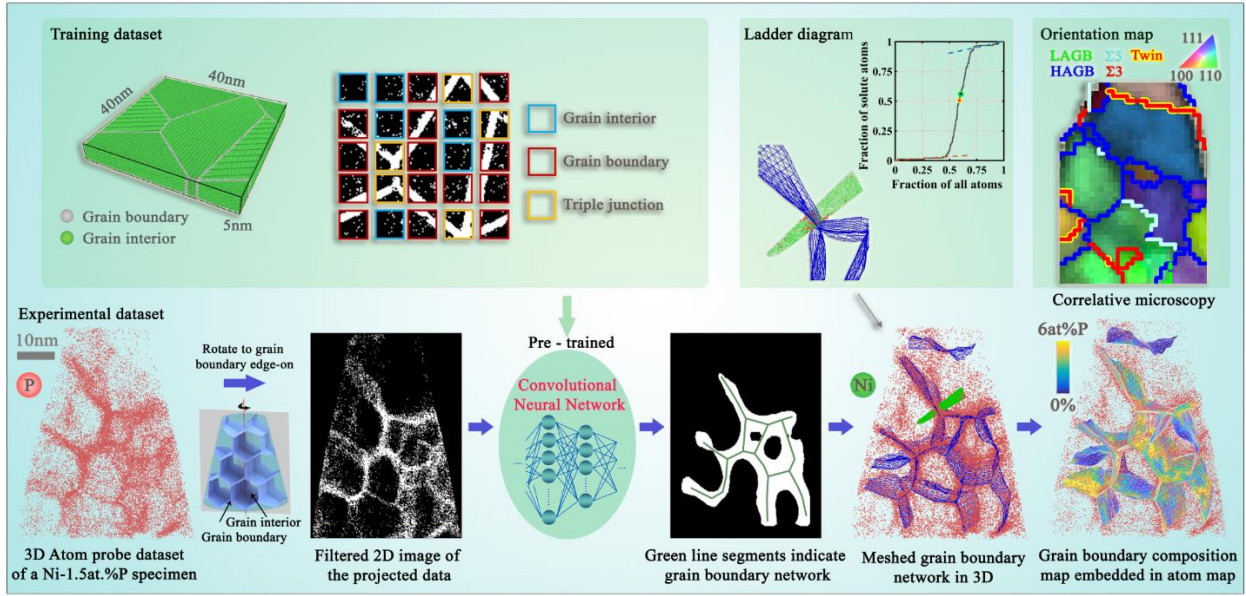

# 1. Introduction

Grain boundaries (GBs) are interfaces between adjacent grains in polycrystalline materials. As ubiquitous lattice defects, they govern many properties of materials, such as strength, ductility, wear resistance, conductivity, etc. [1-3]. Their structure and structure-property relationships have always been a subject of intense research [4-14], but the fact that GBs are buried within the material volume does make it more challenging to probe their structures and composition. In an alloy, the reduction in interface energy drives solutes to segregate to GBs, a thermodynamic effect that has been described by the adsorption isotherm [15-21]. Such solute segregation to GBs can leverage a plethora of effects. In some cases, the solute(s) can lead to possible de-cohesion or formation of galvanic elements [22, 23], both of which are deleterious effects acting on strength and longevity, but some solutes can also increase interfacial coherence [24]. In other cases, solute partitioning to the GB can decrease interfacial mobility [16, 17, 25] as well as reduce the excess interfacial energy, a topic that has been key to nanocrystalline stability [16, 17, 26-28]. In addition, solute segregation in GBs can trigger precipitation of secondary phases and influence phenomena such as deformation [29-31], recrystallization [32], and grain growth [17, 33]. These examples how solute segregation to GBs offers multiple opportunities for alloy and microstructure design [29, 34].

The theoretical study of interfacial segregation started a century ago by J. W. Gibbs who treated the interface as an infinitely thin layer with "phase-like" properties [18]. In the Gibbs' adsorption isotherm, the solutes are frequently approximated as species without chemical interactions, i.e. in its original form the Gibbs adsorption isotherm applies to the dilute limit of equilibrium segregation. This means that a decorated GB is assumed to exhibit a uniform in-plane chemical distribution of the segregated specie(s). Later Fowler and Guggenheim [19], Hart [20], and Guttmann [21] investigated interface segregation beyond this dilute limit. Once solutes segregate to an interface, they will most likely not behave as a statistical solid-solution but experience some form of preferred interactions. The rationale behind this type of segregation interaction comes from the observation that segregation increases if the bulk solubility decreases, i.e. if solutes prefer to get trapped at a defect rather than being solved in a bulk solid solution [35-37]. As a consequence, the solutes will also, most likely, assume preferred neighborhood configurations once the solute decorates the GBs [38, 39]. This interaction among segregated species can lead to phenomena such as in-plane GB spinodal-type decomposition [40, 41], GB phase transformations [20], and solute wetting transitions at the GBs [42]. Factors such as solute-solute interaction [19] and local atomistic structure [43-45] all play significant roles in governing the spatial distribution of solute atoms inside the GB planes. Therefore, it is important to understand and quantify the in-plane arrangements that solute atoms at planar defects can assume directly form experimental data, from the dilute limit to complex low-dimensional decomposition patterns (hereafter all referred to as chemical features). To avoid intrinsic bias, methods that quantitatively analyze those chemical features in an automated characterization approach increases the opportunity to discover new interface thermodynamics features, kinetic phenomena, and the associated chemistry-structure-property relationships which had remained elusive for so long.

Techniques to characterize the solute composition at GBs require high spatial resolution and high chemical sensitivity. Probing approaches for this purpose include transmission electron microscopy (TEM)-based techniques, i.e. energy dispersive X-ray (EDX) and electron energy loss spectroscopy (EELS) [46-49], as well as atom probe tomography (APT) [6, 14, 50-56]. TEM-based methods usually provide a two-dimensional (2D) projection of the solutes in the GBs [57]. Thus, in such cases, it is difficult to distinguish whether the solute atoms segregate homogeneously, i.e. occur as a dilute coverage without chemical spatial correlations, or form a discontinuous pattern along a GB plane driven by element-specific chemical interactions. In contrast, APT provides a three-dimensional (3D) characterization of GBs and can accurately quantify solute distribution on the GB plane [13, 14, 50]. Thus, the 3D reconstruction from APT offers one additional dimension of information to TEM-based techniques, when exploring chemical features in GBs. Typical visualization tools for APT include chemical iso-composition or iso-density surfaces, which help to highlight the location of features with a specific composition or density value [58-60]. One-dimensional (1D) line profiles and 2D contour maps can aid to quantitatively displaying of the local composition [61]. In the specific case when consistent composition gradient directions can be computed for a region in the APT dataset or when the iso-surface patch culls a closed polyhedron, it is possible to compute signed composition profiles. Such a profile, centered at the interface and aligned with a consistent outer or inner unit normal, respectively, are known in the APT community as a proximity histogram (or "proxigram" for short) [62]. These methods can be used to reveal various aspects of local chemical compositions at defects. However, there are still challenges when it comes to the quantification of chemical features along a GB plane that is arbitrarily located and inclined in 3D space and carries complex chemical patterns. For example, interfaces in polycrystals may not be flat but exhibit topological variation for a variety of reasons including equilibrium of forces at nodes and/or interfacial energy minimization [9-11]. Such complex structures of interfaces make the mapping process in APT arduous and potentially user-biased. Below we summarize key studies that have tackled with problem in how to identify and map solutes in APT reconstructed GBs. [6, 54-56, 63-66].

Yao *et al.* calculated the solute distribution map from a curved GB plane based on the Hough transformation of reconstructed APT data sets [6, 63]. With the assistance of supervised machine learning (ML) algorithms, referred to as boosting [67], Wei *et al.* were able to characterize the five kinematic degrees of freedom that define a GB (misorientation and plane inclination) with near-atomic resolution [68]. This method enabled the direct extraction of local compositional and geometric information (i.e. the curvature of the interface) without human-perceptive bias from manual manipulation procedures or user experience [68].

Nevertheless, these two solutions are not sufficiently robust when descriptive crystallographic information about the adjacent grains is missing [6, 63, 68, 69]. In other cases, the segregated solute atoms can serve as markers to locate the position of the interfaces for creating a solute distribution map. Felfer *et al.* used such iso-composition surfaces to determine the location of these interfaces [55, 64, 65]. This method is practical if iso-composition surfaces form a continuous interface set. If the segregation and depletion is not continuous, such as in cases for interfacial spinodals [39-41, 70], spatial gaps appear and the interface is no longer uniform (complete) in this visual representation. Furthermore, calculating iso-composition surfaces with implementations of the marching cubes algorithm do not necessarily produce topologically correct results, nor can do they guarantee a description of closed surfaces [71]. In these cases, where discontinuities exist, Felfer *et al.* reported that the interface needs to be manually located, making the analysis user-dependent, with uncertainties that are difficult to quantify and limited reproducibility when the same data set is examined by different researchers. In a recent publication, Peng *et al.* employed a principal component analysis (PCA) to locate and reconstruct solute decorated interfaces [54]. This method offered a more reproducible alternative for locating the position of the interfaces. However, its reliability is not guaranteed when multiple grains exist simultaneously in one APT data set. In all of these pioneering studies, the most critical challenge associated with mapping the chemical decoration on the interface is the ability to locate the interface in an accurate and reproducible manner.

Here, we propose a ML framework that unbiasedly predicts the location of GBs from an APT data set for local composition mapping [72]. The choice of ML algorithms over traditional image processing methods is motivated by several concerns. In traditional image analysis approaches, several drawbacks exist that include the following: high expertise and manual training is normally required with no automation; poor reproducibility can occur between different datasets; analysis often consists of time-consuming workflows; when automation options are available, they usually still require some partial manual data processing; and, finally, insufficient sensitivity for pattern recognition by human users as compared to well-trained machines. For example, the Hough transform [73] is a traditional image processing method that uses a predefined criterion to identify linear objects in images. In cases where complex patterns exist, some features maybe 'hidden' from the user's direct observation unless they have some *a priori* understanding or knowledge. To avoid this user-defined issue, we choose convolutional neural networks (CNNs) [72], a core method in deep learning for complex image recognition. From a methods perspective, CNNs are multi-layer artificial neural networks with backpropagation and adequate penalizing functions (considering both, penalization of total prediction deviation and of over-fitting) where the individual layers are usually specialized on feature-specific filtering and compression operations. These are usually implemented in the form of marching matrix dot product operations that are applied in a staggered sequence to the input layer data. The weight functions behind the filters are iterated by training, as for all neural networks. CNNs are usually strong in the automated recognition and the classification of the features formed by complex patterns. After adequate training, CNNs can recognize different types of objects in complex patterns [74, 75]. Such an automated ML approach can accelerate the analysis process and improve reproducibility. CNNs have the potential, in some cases, to reveal patterns that remain hidden when analyzed by using traditional methods or when using forward pattern analysis with consideration of only known pattern features. To this extent, some pattern features can likely be identified more efficiently and in less biased form by the user when using ML algorithms. ML has proven capable of such advantages, specifically in case of APT data analysis, as shown by the papers of Wei *et al.* [68, 76], Madireddy *et al.* [77], and Zelenty *et al.* [78].

This paper expands upon these methods by reporting a new workflow for an automated approach to quantify the spatial distribution of solute segregation along GB planes from APT data sets. In this method, decorated species assists in determining the positions of the GBs. Specifically, we show how to train a CNN [72, 74, 75] with synthesized images for recognizing and labeling local features, such as images showing regions-of-interest probing grain interiors, GBs, and triple junctions, from APT data sets. Once the location of the GBs is identified, the quantification of the solute distribution along the GB planes follows the framework introduced by Felfer *et al.* [55, 64]. In particular, this work tackles the step of how to automatically and accurately determine the positions of GBs where mapping is needed. The ability identify GBs in an unbiased manner via automated methods is an essential need for consistent APT analysis, particularly for APT datasets which contain multiple GBs. We have successfully employed this ML-based approach to study the GB solute segregation phenomena in three cases, namely (1) Ni-P, (2) Pt-Au, and (3) Al-Zn-Mg-Cu alloys. Such a quantitative, unbiased, and automated characterization method is capable of catalyzing new discoveries related to interface thermodynamics, kinetics, and the associated chemistry-structure-property relations.

## 2. Results and Discussion

### 2.1 Workflow for the in-plane grain boundary composition analysis approach

Fig. 1 shows the workflow for determining the in-plane GB composition. As a first test scenario, we selected different polycrystalline specimens with near-columnar grain morphologies. The advantage of a columnar grain is that the planar interface is relatively flat and parallel in one direction readily enabling the GB composition to be directly captured in a 2D plane [79]. Below we summarize our approach followed by more detailed sub-sections for each step.

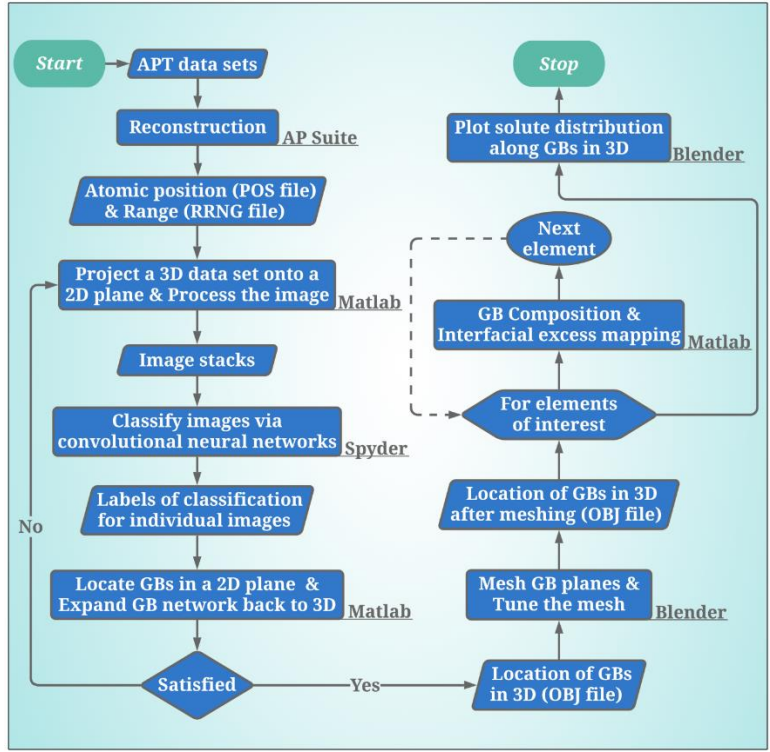

**Figure 1** Flow chart summarizing the steps of ML-enhanced mapping of GB composition and interfacial excess from APT data sets. The term 'AP Suite' refers to the atom prober's toolkit for data analysis workstations [80]. Matlab is a numerical computing environment developed by MathWorks, Inc. Spyder is an open-source integrated development environment (IDE) for scientific programming in Python. Blender is an open-source computer graphics, animation, and compositing software [81]. POS is an APT file format for storing atomic positions and associated mass-to-charge-state ratios. RRNG is a range file format identifying the chemical information of each ion species in APT data by associating an element or molecular ion with a set of mass-to-charge-state ratios ranges [82]. OBJ is a geometry definition file format. 2D and 3D refer to two- and three dimensions, respectively.

First, we start with the APT data set, which contains two types of information: (1) a file for the positions (x, y, z) and the respective mass-to-charge-state (m/q) ratio for each ion to that position and (2) a m/q range file which identifies each ion species by associating a set of mass-to-charge-state ratio intervals, or a complex of ions (molecular ion), that occurs from the uncertainty associated with the arrival time variance upon detection. Such data sets can be experimentally measured or computationally generated. We used the workstation (AP Suite 6.1) to perform the initial data reconstruction on the experimentally captured data set [82] and applied an in-house developed Matlab (2018a) script to generate the synthesized dataset [83]. Both data sets are readable in AP Suite 6.1 for extracting and exporting regions-of-interest (ROIs) as the input of the subsequent GB compositional mapping and analysis. For this work, our ROI was cylindrical because of the symmetry in the reconstruction associated with the composition within the GB plane.

In the next step, we rotate these datasets to calculate successive projections of the solute composition onto the planes to identify specific rotation angles of so-called edge-on configurations where the grain column axis is perpendicular to the current projection plane. Section 2.2 contains more details about this particular step.

In section 2.3, we used a trained CNN to identify where the compositional signal strength is sufficiently above background to support the spatial position of the GB network on the projection plane. This results in a skeleton network of the chemical decoration, quantified here via an average solute composition projected onto the plane. Given the constraint that we analyze columnar grains, we use this skeleton to extend the GB planes in 3D. We developed an algorithm for meshing these GB planes and tuning their morphology to track the GBs in 3D using in the open-source software Blender (version 2.92) [81]. Section 2.4 explains the quantification of the GB segregation, with a workflow step that has been coded in Matlab (version 2018a). Section 2.5 validates our approach with a synthesized data set. Finally, in section 2.6, we applied the approach to analyze the in-plane GB chemical features from experimentally obtained APT data sets from three different alloys to assess the robustness of our workflow and proposed methods.

## 2.2 Projection of three-dimensional chemical information onto a two-dimensional plane

Our method has first been applied to APT data retrieved from a ZnO-Al thin film. An evident columnar grain morphology was present with strong segregation of aluminum (Al) to these GBs, as shown in Fig. 2. Because of this strong partitioning, we can readily identify the GBs in an edge-on configuration, where the projection plane has been aligned perpendicular to the cylinder shaped GBs, see Fig. 2A.

In this step, we detail an algorithm to automate the detection of the GB direction that is orthogonal to the GB normal, or, in other words, the direction that is parallel to the GB plane. This method mainly focuses on the interpretation of the chemical feature of GBs, i.e. GB composition and the corresponding edge-on and in-plane patterns. Note that atomistic structures, i.e. local atomic motifs in the GB, are not resolvable because of the spatial reconstruction limitations and detection efficiencies associated with APT. We assume that the chemical decoration is substantially higher in the proximity of the GBs than in the grain interior. In this case, we expect that the chemical contrast forms a connected trace image of the GB network, instead of only a set of isolated traces for arbitrary rotations of the dataset.

Averaging the 3D chemical decoration along these aligned grains allows for a clear distinction between the grain interior and the GBs, owing to the above solute segregation. To determine the specific projection plane, we generated a series of 2D projection images by successively rotating the data set by an angle $\theta$ about an axis which was perpendicular to the grains' columnar axis. Fig. 2A presents a schematic diagram to show this rotation. The specific axis chosen here coincides with the field evaporation axis that was used in the APT experiment. Choosing this axis has the advantage of minimized APT reconstruction aberration effects [60, 84]. Thus, at a certain rotation angle, $\theta$, a GB edge-on configuration is achieved and a clear edge-on GB decoration projection is now present with minimal aberrations in the reconstruction.

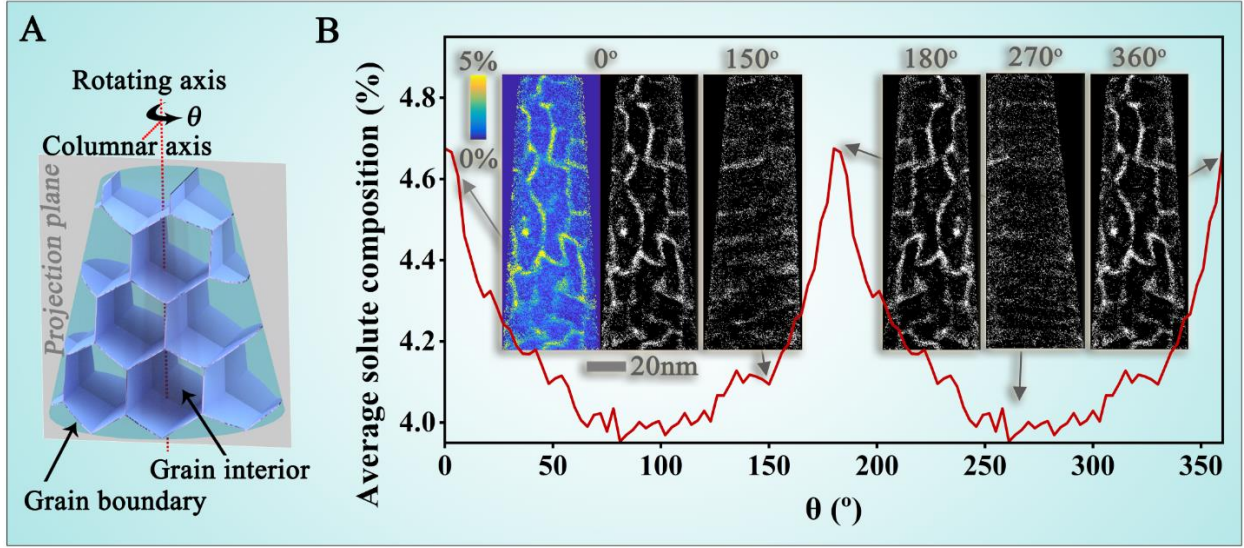

**Figure 2 A** Schematic of an atom probe tomography specimen showing the rotation by an angle $\theta$ about an axis perpendicular to the grain columnar axis **B** Average solute (aluminum) composition as a function of $\theta$ in the filtered 2D image of the projected APT data obtained from a ZnO-Al sample with columnar grain shape. The embedded composition map exemplifies the average (projected) solute distribution in the projection plane, as shown in the left-image at 0°. We also include five filtered 2D images at several $\theta$ values (right-image at 0° and images at 150°, 180°, 270°, and 360°, repectively) for comparison. The average solute composition is calculated from regions highlighted by white pixels in the filtered 2D images.

A criterion is now defined for rendering this formally, i.e., for reaching an optimum $\theta$ value. First, we binarized the composition map by filtering the non-random solute distribution via a threshold $T$ defined as

$$T = \mu + k \times \sigma \tag{1}$$

where $\mu$ corresponds to the mean composition for all pixels in the projected composition map, $\sigma$ is the standard deviation among all pixels, and $k$ is an adjustable number. Regions with compositions higher than the threshold $T$ were assigned a "1", and regions having compositions below that value are "0". The filtered pixels, for which values are "1", correspond to regions enriched in solutes (the projection of the chemical decoration). With a GB in an edge-on configuration, the solutes along the GB planes reside within a set of thin line segments in this projection plane, and the average solute concentration, indicated by the filtered pixels, tends to be high on these lines. Using this filter, the GB recognition process will be less sensitive in locating composition variations, and it is adaptable to various composition ranges. While this threshold does not enable accurate compositional analysis in the GB, evident by the binary "1" or "0" binning, its purpose is to locate and define the GB by the presence of solute segregation, not necessarily the quantification of solute segregation. This latter measurement is done using another method used later on in our workflow, Fig. 1.

We now illustrate this process with the ZnO-Al experimental data set. The 3D chemical information was projected successively on multiple 2D projection planes, Fig. 2B. For example, the left-image is at 0° and shows a projected composition map. We binarized the composition map using the aforementioned binary threshold approach, equation (1), and represent the threshold outcome as a black ("0") and white ("1") contrast image, Fig. 2B. Here, $T$ is set at 2.3 at.% when the adjustable $k$ is of 0.5. As such, the average solute composition within the white pixels was plotted as a function of the rotation angle, $\theta$, in Fig. 2B. When $\theta$ is 0°, 180°, and 360°, the maps appear to have some interconnected line segments, which is the signature of a projected aggregate of edge-on oriented prisms (GB network). For those $\theta$s, the average solute composition tends to be the highest. Fig. 2B also includes the maps at $\theta$s with a low average solute composition, such as 150° and 270°, where the parallel lines run through the image. Note that GBs oriented in this direction are not suitable for the later auto-detection of the GB. Therefore, we chose the $\theta$ value to be 0° for an edge-on GB configuration.

### 2.3  Automated grain boundary detection by using a convolutional neural network

The convolutional neural network comprises multiple layers with the goal of building a function to classify an input image [72, 85-87]. The individual layers are specialized for feature-specific filtering and compression. For instance, the convolutional layer extracts structural features, e.g. an edge, a line, or junctions, from the source image and passes its results to the next layer [87]. This convolutional layer helps with sharpening, blurring, noise reduction, edge detection, or other operations that can assist the ML algorithm in learning the specific characteristics of an image. The weight and bias functions behind the filters are iterated through training [72, 87]. For convenience, we refer to this automated GB detection by a CNN as Auto-GB-CNN.

#### 2.3.1 Training data of synthesized grain boundaries for recognizing local features

The preparation of training data to recognize actual data is a crucial step for accurately capturing features of interest in any supervised ML algorithm. With a sufficiently large set of training data, CNNs can efficiently learn the weight factors pertaining to all filters and provide a translation equivariant response. This is achieved by identifying adequate penalty measures from the training data set. The second important aspect relates to the avoidance of overfitting. Even if some experimental features, i.e. chemical contrasts that form a coherent trace image of the GB network, are not contained in the training data, ML can identify them by capturing features of interest from the learned structures. CNNs have a considerable potential to unbiasedly and automatically reveal chemical patterns that are hidden when using forward pattern analysis with consideration of only known pattern features.

Our goal is then to create training datasets containing grains with different shapes and GBs with varying degrees of solute decoration for mimicking compositional information obtained from experimental APT data sets. To do this, we created various simulated data sets with different (but known) compositional information and GB structures. This was achieved by generating 2D Voronoi tessellation and then extending its trace into 3D to mimic the columnar grain shapes [83]. Using these simulated structures, we then filled the columnar volumes with periodically positioned atoms defining different orientations per column to generate various GB misorientations and inclinations. We did not relax the atom positions of these simulated structures for the following two reasons:(1) The spatial resolution of APT is approximately 0.1-0.3 nm in depth and 0.3-0.5 nm laterally [61, 82, 88] with field evaporation of ions near GBs being further displaced in the reconstruction from local magnification effects [60, 84, 89]. Therefore, it is beyond the capability of the experimental APT technique to capture the atomistic structure of GBs precisely and accurately. Hence, any simulation relaxation to further improve the atomic position for training would not ultimately improve the ability of detecting such features when applied to the experimental data set. (2) We will use the atomistic structures only to distinguish the chemical difference between the grain interiors and the GBs. The actual arrangement of atoms at GBs is of less interest to this particular work. Fig. 3A is a simulated 3D image of a polycrystal with columnar grains. The grains are defined as distinct regions of atoms (represented as green dots in Fig. 3A) which are arranged on positions of a face-centered cubic (FCC) crystal lattice with a lattice parameter of 0.4 nm. The remaining grey-colored atoms within the simulation are atom positions located in the GBs.

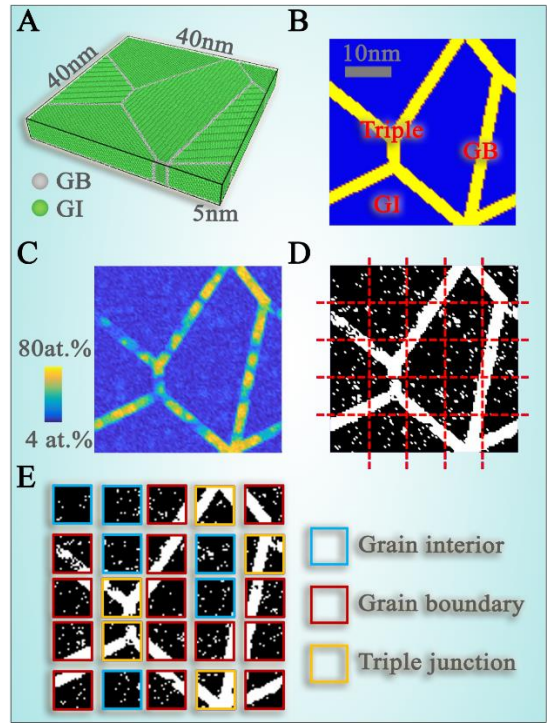

**Figure 3** Synthesized training data set. **A** 3D atomistic structure of columnar grains, where atoms in grain interiors (GIs) are green, and those at grain interiors (GBs) are grey. **B** The projected 2D map showing resolved GIs, GBs, and triple junctions (Triple). **C** The projected 2D composition map after adding solute atoms as a form of clusters. **D** The filtered 2D image is divided into small sub-images by red dashed lines. **E** We assign each of the sub-images into three groups by the different colors of their frames.

Note that local magnification effects [60, 84, 89] mentioned above will cause the APT reconstructed GBs widths to potentially appear thicker than their actual width. To account for this effect, we defined artificial GB widths over which resolved segregations would occur to 3 nm, which matches more closely with anticipated experimental APT observations. Fig. 3B is a representative projected 2D image showing the resolved grain interiors, GBs, and triple junctions. Since compositional variations can occur both in the grain interior and along the GB planes, we also added solute atoms to interior and GB regions as clusters with a range of sizes, various compositions, and different number densities. Fig. 3C is a 2D projection of the 3D simulation with these additions for an average composition of 5 at. % in the interior and up to 80 at. % enrichment to the GBs. These modifications also follow the experimental findings, namely that GB segregation is typically a phenomenon where in-plane GB compositions can be highly inhomogeneous and vary by more than an order of magnitude [29].

The actual compositions of grain interior and GBs are related to the alloy system, solute composition, crystal structure, and heat treatment. Here the solvent and solute elements are not defined, as this training data set had been designed for an arbitrary alloy system with GB segregation. We have generated a range of compositional combinations between the grain interiors (e.g. 3 - 6 at. %) and GBs (e.g. 2 - 80 at. %) to cover a wide range of potential possibilities. Using the filter from Eq. (1), the simulated GB network can be well revealed in the binarized 2D image, Fig. 3D. Sub-images from Fig. 3D, were then divided, shown by the red lines, with these sub-images used as the training data sets. Each of these images, delineated by a colored box in Fig. 3E, is classified into one of three categories – grain interiors, GBs, and triple junctions. The criteria for distinguishing which category the image belongs to are based on how many grains the image contains. For example, we define that an image that involves only a single grain belongs to the grain interior category of images. If two grains are present, the image is labeled as a GB and all other identified features then belong to the triple junction category. Using the above method, we generated the training data set with 20,000 images of the same size but different patterns. Learning to back out compositional features, such as GB network traces (lines) and triple-line-imaging plane intersection points, can then help the CNNs locate the GB network from the experimental APT data. The following section now discusses the details of this classification process by CNNs.

**2.3.2 Convolutional neural network for recognizing grain boundaries**

The CNN used herein is implemented in the software library TensorFlow 2.0.0-beta1 [90]. The input data of a CNN has a shape of (number of images) × (image width) × (image height) × (image depth). In this work, the input data is a (20,000 × 20 × 20 × 1) matrix with each input image having a 20-pixel width, a 20-pixel height, and a 1-channel depth. The edge length of each square pixel represents 0.4 nm spatial distance. The training images are composed of grains with different shapes and various solute distributions. The left image in Figure 4 is an example of an input image with which the CNN has been trained to detect it as a triple junction unit. Fig. 4 also contains the neural network schematic for the training process.

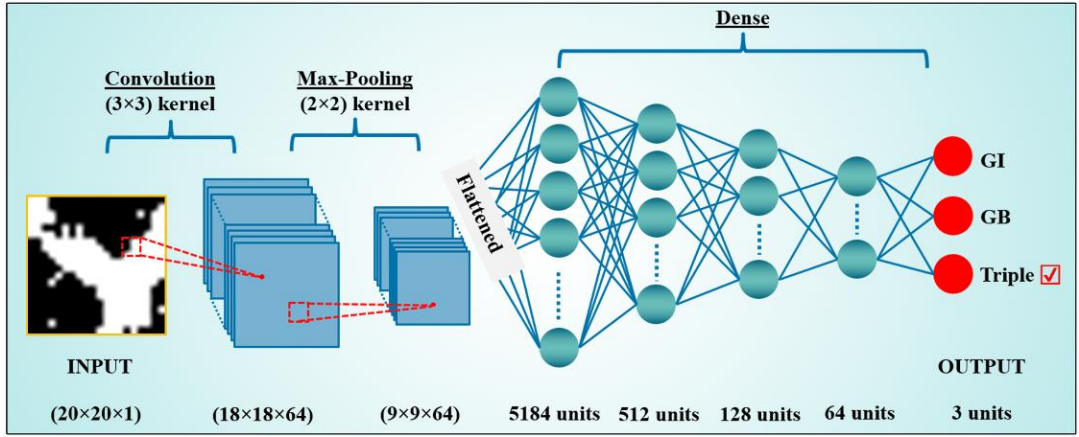

**Figure 4** Construction of the convolutional neural network for recognizing local features. Here, the abbreviations GI, GB, and Triple stand for grain interior, grain boundary, and triple junction, respectively.

A convolutional layer is composed of a series of filters known as kernels, namely a matrix of numbers. In our work, the convolutional layer utilized a $3 \times 3$ pixel$^2$ kernel whereupon we move the sampling window (i.e., the field of view for the kernel) over the input image, pixel after pixel, for every position in the input image. Hence, the kernels only "see" a portion of the input image. The dot product of the imaging matrix with the operations of the kernels generate a single value to present the analyzed pixels. By marching the sampling window through the input image, and performing this dot product operation, a feature map is created. We used the Rectified Linear Units (ReLUs) activation function to transfer the dot product value into the next layer, which outputs the inputs directly if it is positive value but if the operation yields either zero or a negative value, the output is assigned a 'zero' value [87]. As a default activation function for the convolutional layer, ReLU can train the model more easily while achieving a better performance. For example, it can overcome the vanishing gradient problem that frequently occurs with sigmoid and hyperbolic tangent activation functions [87]. The output after the convolution layer retained the same number of images, but each image now becomes a matrix of $(18 \times 18 \times 64)$. Here, 18 pixels refer to the width and height of an image, while 64 is the number of channels.

By adding a layer of $2 \times 2$ pixel$^2$, the largest element compression kernels reduce the number of parameters while retaining information in latent space [87]. This is known as a max-pooling layer. During this step, the kernel was stridden over the input matrix by moving it horizontally every two pixels column-wise and vertically every two pixels to subsequent rows. This operation provided an output feature map with the value extracted from each kernel. The height and width of the image decreased in half forming a compressed matrix of $(9 \times 9 \times 64)$.

Afterwards, we flatten those output layers into different classes. This step transforms the two-dimensional matrix of features into a vector that can be fed into a fully-connected neural network classifier. In our case, the first fully-connected layer contains 5,184 units after flattening. We further condensed sequentially all units into 512, 128, and 64 units. In the last pooling step, the activation function, Softmax, normalized the output of the network to a probability distribution over the predicted output classes [87]. As such, the outcome of the neutral network classifier is a label representing one of three classes (grain interior, GB, and triple junction) for each image.

We have tested multiple parameters, i.e. the number of convolutional layers (1-3), the number of channels (32, 64, 128), and the size of dense layers (512, 128, 64). The neural network shown in Fig. 4 gives the highest accuracy and was therefore selected for the later recognition step.

We now discuss the influence of the number of training images on the accuracy of the CNN as well as potential inaccuracies in our proposed ML approach. We trained the CNN with a different number of images and then calculated the sparse categorical accuracy, i.e. a value indicating how often predictions match integer labels [90]. The calculations were based on the same 1,000 test images. Fig. 5A shows the sparse categorical accuracy of the CNN as a function of the number of training images up to 20,000. Each point in the plot is an average of five different groups of training images, with the error bars showing the variance. The accuracy rapidly increases as the number of training images increases. At approximately 12,000 training images, the accuracy improvement levels out at a value of 0.87 approximately. Fig. 5B shows the confusion (error) matrix between the predictions and the actual values. Distinguishing between the grain interior and the triple junction was relatively accurate; however, most of the ambiguity or error occurred when the system needed to distinguish between either a GB and triple junction or a GB and a grain interior. These ambiguities are a result of the following: Firstly, if there is only a small segment of a second GB connected to a triple junction, the CNN can mislabel it as a single GB. Secondly, since clusters can also exist in grain interiors, there is an uncertainty in distinguishing this type of grain interior from such GBs that only carry a low amount of solute segregation.

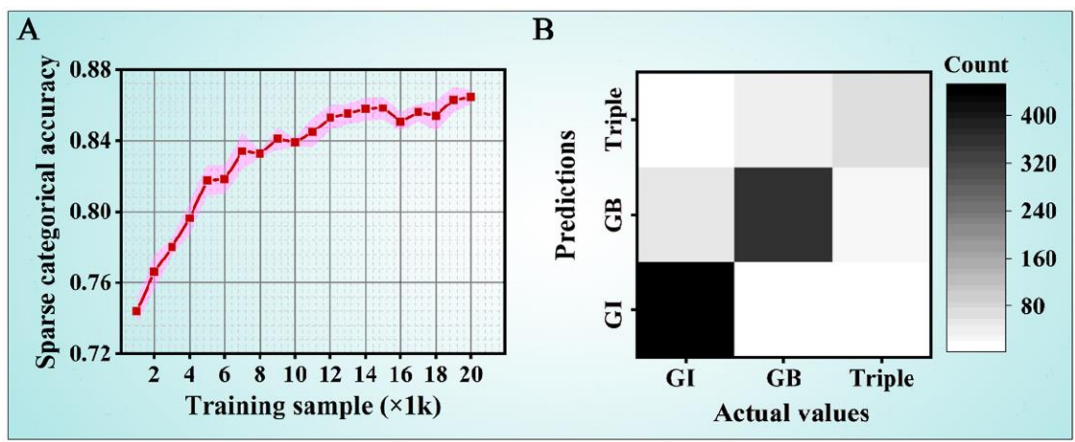

**Figure 5 A** Sparse categorical accuracy as a function of the size of training samples. **B** Confusion (error) matrix between predictions and actual values. GI, GB, and Triple represent grain interior, grain boundary, and triple junction, respectively.

While these ambiguities are concerning, they can be mitigated by employing a weight and bias function that can automatically recognize GBs by training it as well with the 20,000 images. Here, we divided the experimental image into smaller field of views or windows that were 20 × 20 pixel² whereupon the feature (grain interior, GB, or triple junction) in the window is automatically detected and identified by the above CNN algorithm. By overlapping the windows, the continuity of the feature is better identified. Through this window scanning approach, some degree of misclassification that occurs will not significantly alter the correct detection of the location of GBs. The actual results relate to what the region is most likely identified as. In other words, even if a region is not accurately indexed in one window but correctly identified in the surrounding windows, the probability to correctly recognize this region is still quite high. We have also used this image processing to remove artifacts generated during the CNN step. In the next paragraph, we provide the details of this image processing by using an example of the Auto-GB-CNN process on the Ni-P APT dataset [91].

### 2.3.3 Example of automatic grain boundary detection from atom probe data sets without correlative microscopy

Here, we apply the methods discussed in Sections 2.2 and 2.3 (Auto-GB-CNN) to locate the GB network in a columnar structured Ni-P APT data set where the phosphorus (P) has partitioned to the GBs [91]. Fig. 6A shows the filtered 2D image of the averaged solute composition with most of the GBs being in an edge-on configuration. This 2D image is composed of $263 \times 223$ pixels². Using a sliding window of $20 \times 20$ pixels², the CNN examined the composition-related features in each window and classified it into one of three categories (grain interior, GB, and triple junction). When the sliding window was labelled as grain interior, we set all pixels to "0". For images of all of the other labels (GB and triple junction), these pixels are now set to "1" and stored separately. Afterwards, we superimposed the assigned values in all the sliding windows into a series of new 2D matrices that have the same size as the original 2D image. Those two new matrices represent a GB map (Fig. 6B) and a triple junction map (Fig. 6C), respectively. In such a manner, features, i.e. GBs and triple junctions, have been counted multiple times during the sliding window step. Consequently, these features result in a high detected value in the output images. In Fig. 6C, we also highlight the local maxima in these distinct regions, which could be the location of possible triple junctions, by green spot makers in the image. Fig. 6D shows the binarized GB signal as a black and white image by using the same filter described in Eq. (1). We then applied the "skeleton" algorithm [92] to reduce all objects to single-pixel width lines. As shown in Fig. 6D, the line segments created in this way (colored in red) appear as an interconnected GB network. We further divided them into multiple line segments by locating the ending pixel or junction pixels. Here, each line segment represents an individual GB. The GB network detected by the GB feature is not always perfect, i.e. the red edge lines in Fig. 6E do not match well with the composition map. This issue arises from either the inaccurate prediction of the CNN or a 'blur' due to the sliding window process. We can resolve this issue by simply adjusting the *k* in Eq. (1) to make the binary image sharper. As a second option, the position of the GB line segments can be adjusted with the help of the junction pixels detected by the triple junction features. Fig. 6F indicates modified GB line segments (green edge lines) with all GB line segments extended along the columnar direction to form the interconnected GB planes. This operation involves a translation of the coordinates of the GB lines. Finally, we exported the 3D geometry of the GB network into a geometry definition file format (OBJ file). This file is a standard 3D image format that can be opened by various 3D modeling programs.

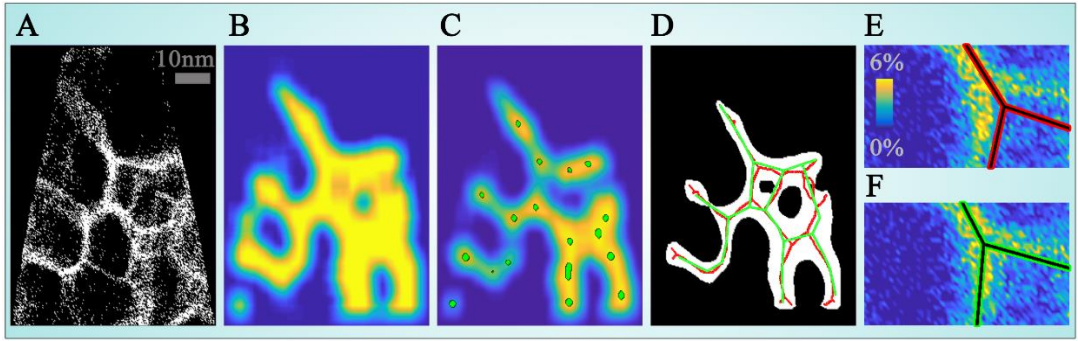

**Figure 6 A** Filtered 2D image from the experimental Ni-P atom probe data set [91]. **B** Pixels in yellow highlight the GB regions as they were detected by the CNN. **C** Pixels in yellow represent the triple junction regions as they were detected by the CNN. Green-colored spots highlight the possible triple junctions. **D** The black and white image is the binarized image of B according to the filter described in Eq. (1). Here, the red line segments are the GBs detected by the GB features. The green line segments show the modified GB network favored by the triple junction features. Overlaid onto the composition map are GB line vectors before (red edge line in **E**) and after (green edge line in **F**) being adjusted by the triple junction criterion.

### 2.3.4 Meshing of the grain boundary planes in three dimensions

We imported the extracted network of GB traces and triple junction locations into the open-source 3D graphics suite Blender (version 2.92) [81] to create a mesh and perform a local refinement of the mesh. Fig. 7A shows a section of the 3D geometry of the GB network that is overlaid onto a rendering of the P atom positions within the reconstructed Ni-P APT data set. In Blender, it is possible to adjust specific nodes of the 3D geometry to achieve a better match between the highlight planes and the GB positions. For instance, we could manually add a GB that has been confirmed by complementary TEM results but remained undetected by the ML algorithm. A possible reason when this would be needed is for cases when the solute segregation is too low and remains below the threshold to trigger the insertion of a GB segment in the Auto-GB-CNN step.

The next step is to mesh the GB planes. As it is often difficult to obtain meshes with as closely as possible equilateral triangles GB planes, we first cut the GB planes into small rectangles and then divided each rectangle into two triangles in Blender [81]. The minimum side length of the triangles is adjustable for controlling the density of triangles per mesh surface unit area. For Fig. 7B, the minimum meshing length was set to 3 nm. Subsequently, all nodes on the meshes that lie outside the convex hull of the specimen were deleted [93]. Next, we assigned the nodes on the meshes to the respective GB line segments defined in the previous step to identify each GB as an individual unit of the 3D mesh. Finally, in the last step, the positions of the nodes on the GB meshes were automatically adjusted towards the center-of-mass of the solute species using the algorithm introduced by Felfer *et al.* [55]. Fig. 7B is a representative image of the meshed GB network in 3D.

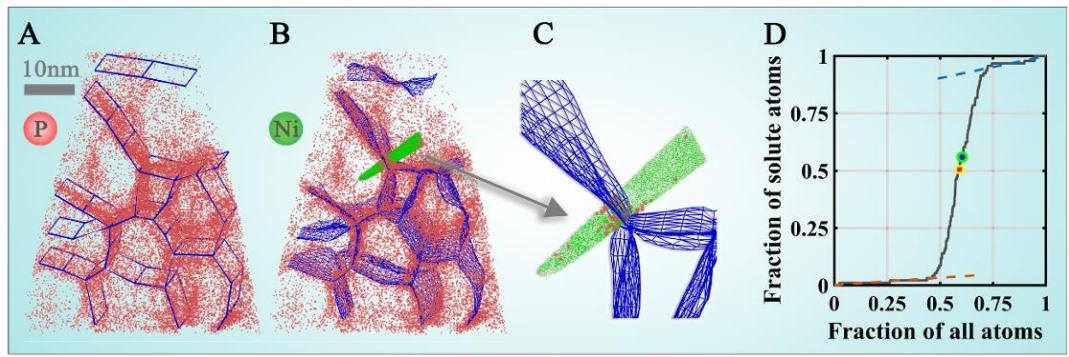

**Figure 7 A** 3D rendering of the P atom point cloud (red-colored) superimposed onto the 3D geometry of the GB network (blue-colored). **B** GB meshes (blue-colored) showing together with selected P (red-colored) and Ni (green-colored) atoms. **C** Magnified region (as pointed by the arrow) which exemplifies a volume used for computing IE and GB composition. **D** The ladder diagram computed from this volume shown in C. The green-edge circle indicates the location of the node in the GB meshes, while the yellow edge circle is the center of the region used for calculating GB composition.

### 2.4 Quantification of grain boundary segregation

With the GBs now defined, the measurement of local solute composition in the GB is merely computing the relative percentage of one specific species among all identified atoms in a given volume. This requires a definition of the GB width. In reality, the crystallographic width of a GB is typically on the size of one to a few atomic layers in thickness [94]. However, the measured APT GB width can span one monolayer to a few nanometers dependent upon how the chemical demarcation is defined in the atom probe reconstruction [95-97]. It can even vary from region-to-region along a singular GB plane depending upon the spatial distribution of the solutes [54]. Consequently, the widths of GBs as measured by APT may appear modestly thicker than their actual width because of these effects that are largely contributed to a 'local magnification effect' when the data is reconstructed [60, 84, 89]. Therefore, the quantification of the precise GB width from APT can be difficult to ascertain. One means to determine the solute content in the GB is the use of interfacial excess (IE) mapping, where IE represents the excess number of atoms per unit area due to the presence of an interface [1, 18]. This quantification is much less sensitive to reconstruction artifacts and does not need one to define the exact width of a GB [54-56, 66]. By combining composition and IE mapping, we can now achieve a more complete description of the solute distributions along the GB planes [60] with our calculation following the methodology from Felfer *et al.* [55]. In summary, the use of the ML-based approach provides a quantitative, less biased, and automated access to identify the GBs whereupon methods of composition analysis on such boundaries can then be readily applied to collectively yield a robust means for chemical characterization from APT datasets on these features. Using this approach, we now continue its application using the Ni-P APT data set.

### 2.4.1 Automatic interfacial excess mapping

Obtaining an IE value requires two quantities: (1) the number of excess atoms and (2) a reference area on the surface of the GB mesh where the calculation is to be performed. We can gain the number of excess atoms from plotting a so-called ladder diagram [56]. Such a diagram is computed for each vertex on the GB mesh (identified by our ML methods) by placing a small columnar volume across the GB mesh and plotting the number of solutes as a function of the number of total atoms when counting both numbers (solutes and total atoms collected) from one end of the volume to the other. When the solutes enrich a GB, the slope of the middle line segment will become steeper than those on either side of the GB. This created the appearance of a ladder shape. Fig. 7C shows the selected volume across the meshed GB with Fig. 7D revealing this ladder step in the Ni-P system. Fig. 7D also includes linear fits for the two segments on either side of the GB, which has been colored in blue and red dashed lines, respectively. These slopes represent the solute presence within the grain interior. The difference in the intercept values of those two lines at the location of the interface is the number of excess atoms in the GB. We employed a Matlab script to search for the turning points where the ladder diagrams could be divided into these three line segments or slopes [98]. While the slope here is steep and positive, indicating solute segregation in the GB, the slope in the middle line segment does not necessarily have to be larger than the slope from either grain. In that case, the GB is depleted of solutes. The ladder diagram can be used to characterize GB enrichment or depletion when the curve can on the one hand be decomposed into three segments that can all be described via linear interpolation and when the values in the middle segment on the other hand are either larger or smaller than those of the other segments representing the situation in the grains on either side of the boundary. While the ML approach does use segregation as the descriptor for identifying GBs, the reconstruction will create a GB plane where parts of the regions on this plane may have little to no segregation. In those cases, where such regions are sampled, there will be no significant change of slope in this type of IE plot. In such regions, excess atom quantities are zero.

The other quantity for calculating an IE value is the reference area of the respective GB mesh where the ladder diagram is evaluated. As the involved volume for IE calculation has a columnar shape in our case study, we can calculate the area, $S$, using the following equation:

$$S = V/H \qquad (2)$$

where $V$ is the volume of the convex hull [93] of the involved atoms and $H$ is the vertical distance measured between the farthest separated atoms in the direction normal to the GB plane. From the original APT dataset, only regions within a few nanometers on either side from the GB mesh are needed for the IE calculation. Hence, we can extract or clip these volumes out of the entire datasets. Using an automatic clipping algorithm from the ML defined GBs, the calculation of IE maps becomes efficient. For the case system shown here for Ni-P, we set the distance to 7 nm on either side of the interface making $H$ 14 nm.

### 2.4.2 Computing the composition within a grain boundary

When one needs to calculate local GB compositions, this is done by computing the fraction of solute atoms over all atoms in a given ROI volume. The accuracy of such a GB composition map depends strongly on the locations of the ROI to the GB position. While the location of GBs determined through CNN are a good estimation, they still represent an approximate position. In some cases, the solute may shift relative to the boundary's actual location based on how the CNN located the boundary mesh. Ladder diagrams are one alternative solution. Here, one can more precisely refine the GB position if one assumes that the solute is located in the middle of the GB. In Fig. 7D a green edge circle is used to indicate the CNN determined position of the boundary. This location appears slightly closer to the grain on the right side. If we now assume that the solute distribution is symmetric along the GB plane, the location of the GB should be located in the middle position between the two inflection points in the ladder diagram. This is indicated by the yellow edge circle in Fig. 7D. A volume centered at this position would then offer a potentially more accurate positioning of the ROI based on the assumptions above. While these spatial differences are relatively minor in our example, it does highlight that a deviation can and does exist and should be accounted for how one may approach an automatic means of determining GB composition.

### 2.5   Validation through a synthesized data set

While the sections above have shown how the Auto-GB-CNN method functions for each step, even using real experimental datasets to highlight its use for specific steps, it has not yet been employed for either a full dataset analysis that contains multiple GBs nor has it been tested to determine its accuracy for compositional mapping. To validate the collective various steps in the Auto-GB-CNN method, it is now tested against a simulated dataset where the composition and features are well defined.

The simulated dataset was created using the same custom developed Matlab script that generated the training datasets (Section 2.3) [83]. Here, the data set contained approximately 6 million atoms on a FCC lattice (a = 0.4 nm) which consisted of a polycrystalline structure with a columnar grain structure that has a known spatial distribution of solute atoms. Here, the average columnar diameter was approximately 34 nm and the average solute composition in the grain interiors was set to 9 at. %. In addition, spherical clusters with radii of 1 – 1.5 nm were added into the grain interiors with such clusters having a solute content of 15 at. %. The GB widths were set to 2.5 nm with the solute segregation in these boundaries threshold at a minimum of 12 at.%. To further mimic experimental data, we allowed the solute atoms to cluster in these GBs with radii of the clusters ranging from 1.1 to 1.9 nm, up to compositions of 94 at.% with this clustering accounting for 82 % of the GB volume. Considering that the size of simulated GB cluster can be larger than the defined GB width, only the region of the cluster within the GB width was used. Every compositional information of the cluster outside this boundary condition was not used.

Fig. 8A shows the atom map of this synthesized data set. The pink atoms represent the solvent atoms. We employed the blue-colored iso-composition surfaces to highlight localized regions with solute compositions above 30 at. % (hence no clusters within the grain interior are seen because they do not reach this threshold). However, clusters within the GBs are noted. In Fig. 8A, this cluster inhomogeneity is clearly visible from the incomplete connections of the iso-composition surfaces in the boundaries. The chemical information of the 3D data set was then rotated and projected onto a 2D plane to satisfy GB edge-on criteria, Fig. 8B, as already discussed in Section 2.2. The green lines in Fig. 8C reveal the ML GB network as it was detected by the Auto-GB-CNN algorithm. Fig. 8D is the local solute composition along these simulated GB planes. The minimum and maximum solute content are 12 at.% and 95 at.%, respectively. While the measured composition is ever so higher than the input value, it does reveal very close matching and confidence in this auto-detection and mapping approach. With the complete Auto-GB-CNN procedure tested against the simulated dataset, we know apply it, in full, to the various experimental datasets discussed at the end of the introduction section of this paper.

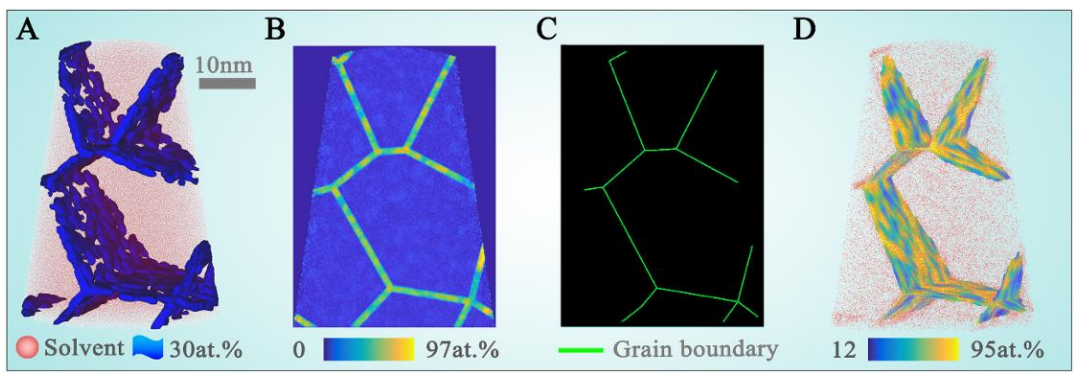

**Figure 8 A** Synthesized data set with 30 at.% solute iso-composition surfaces (blue-colored) embedded. Pink atoms represent any possible solvent atoms. **B** 2D projected image showing local composition **C** Line segments found by the Auto-GB-CNN method profiling the GB network **D** GB composition map embedded in the 3D atom map of solvent atoms (pink-colored).

## 2.6 Applications for experimentally measured atom probe data sets

### 2.6.1 Grain boundary specific segregation

As shown above in the interfacial excess section, P readily partitions to the GBs in Ni. Here we use the Ni-P experimental dataset for further analysis. One of the advantages of the simulated dataset shown above is *a priori* knowledge of the nanostructure features, which is not always available in an APT experimental dataset. To create this *a priori* knowledge to our experimental system, we have identified the GBs in the APT specimen tip, prior to field evaporation, by performing precession electron diffraction (PED) whereupon grain-to-grain misorientation is captured with the data represented by automated crystal orientation mapping (ACOM). Through this type of cross-correlation method, we increase our confidence of the GB identification process by ML in the experimental APT data because the ML generated images can be directly compared to an independent means of microscopy imaging. The details of this cross-correlation method and sample preparation of the Ni-P sample are found in [13, 99].

Fig. 9A shows the PED measurement of the grain orientations. The results document that different GB misorientations are present in the sample, enabling a qualitative characterization of the GB types among low angle GBs (LAGBs, green), high angle GB (HAGBs, blue), as well as a few coincidence site lattice (CSL) boundaries such as Σ5 (cyan), Σ3 (red), and twin boundaries (a specific type of Σ3, red with yellow background). Fig. 9B reveals the reconstruction of the APT data from the ML Auto-GB-CNN method. Nearly all of the GBs were found and matched the PED data. However, the Σ3 twin boundaries were not initially captured by the Auto-GB-CNN method because this boundary revealed minimal solute partitioning. Consequently, these boundaries were added manually to the image shown. This highlights a very important outcome. First, ML may not capture all of the GB features. Hence, using cross-correlative methods that can improve the accuracy of the APT reconstruction is beneficial, as would be the case even if ML was not applied. Even though our method did not capture this twin GB, it readily identified all others and provided for a rapid method of data capture, analysis and representation. This decreases the time for users to sort through the data and apply any manual input to features that are notably absent, which still demonstrates the impact ML can still have on APT reconstruction development. Using the ML method, it was able to automatically reveal the inhomogeneity of P atoms not only between different GBs but also within the same GB, Fig. 9B. This reduces the manual effort and potential human error in such analysis, a major motivation for the ML approach developed here. Fig. 9C displays how GB composition relates to the misorientation. The large error bars notable for all points in this plot indicates significant compositional variation along the GB planes. A more detailed understanding of the distribution of solutes is connected to the potential energy of GB sites and was the subject of another study [91]. This collective study of chemical structure relationships using this ML approach to APT data analysis provides a quantitative and representative way to reveal GB-specific separation phenomena, while also recognizing potential areas for future development in identifying GBs in such datasets where low solute segregation in GBs exists.

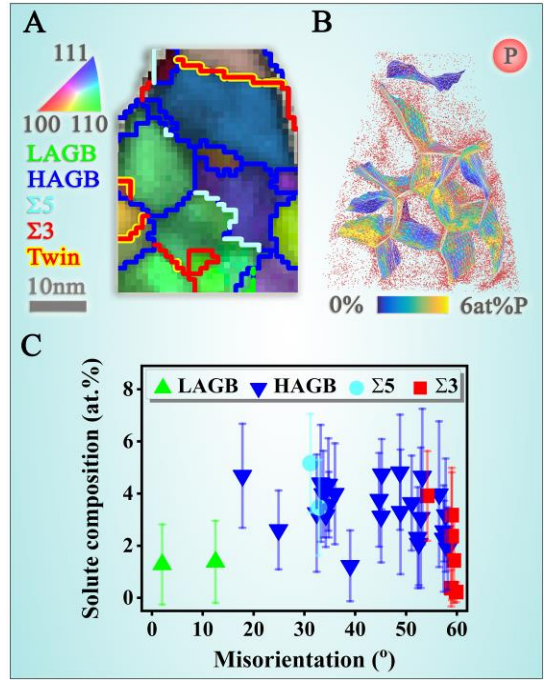

**Figure 9** A correlative study of the 350°C/1 h annealed Ni-1.5at.%P thin film [91]. **A** Precession electron diffraction orientation and GB map for an atom probe specimen. **B** GB composition map embedded in the 3D atom map of P (red-colored) from the same APT specimen. **C** GB composition (P) as a function of GB misorientation.

### 2.6.2 Dislocations inside of a low angle grain boundary

While the majority of the GBs above are associated with large misorientations, LAGBs contain series of periodically arranged dislocations that create modest crystallographic rotations, usually below a few degrees, with the spacing of such dislocations directly linked to the misorientation. Similar to GBs, these dislocation defects also attract solutes. Such solute segregation at a LAGB depends on the local structural disorder and the hydrostatic stress around the containing dislocations [100]. In this second case study, we demonstrate our Auto-GB-CNN approach to reveal the chemical features of these local crystal defects.

In this example, Pt-7at.% Au thin films were sputter-deposited and annealed at 1027°C for 15 minutes to facilitate the solute partitioning to defects. Details of the specimen preparation and analysis are given elsewhere [101]. Fig. 10A and B are the composition map and IE map of the LAGB with a misorientation of 2.5°, which can be further decomposed into a tilt component of 2.43° and a twist component of 0.12°. The LAGB in the APT experimental dataset was captured using our ML methodology, whereupon a pattern of parallel lines was visibly represented in both a compositional and IE map, from the Auto-GB-CNN toolkit above. The lines in this GB are attributed to the dislocation array that creates this misorientation [40, 101-103]. We treated such an arrangement of dislocation as a spatially continuous GB rather than isolated dislocation lines. As for the latter, the excess in the unit of atoms per line can be calculated according to reference [64]. In this work, the compositional variation is revealed over the entire GB plane. Fig. 10C includes a more quantitative result that shows a linear relationship between the IE values and solute compositions, with the data points collected from 910 vertices in this LAGB mesh that was automatically identified by the ML method. By studying the separation behavior along the GB plane in this automated manner, the in-plane chemical patterns that are not nominally easily detected by standard compositional analysis were presented in an unbiased manner.

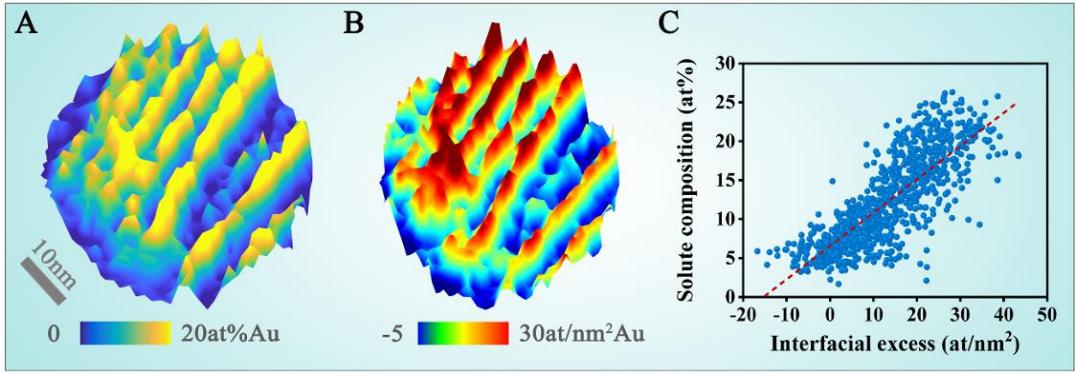

**Figure 10 A** Composition map and **B** IE map of a LAGB found in the APT specimen extracted from a 1027℃ /15 minute annealed Pt-7at.% Au thin film [101]. **C** Scatter plot to reveal the relationship between GB compositions and IE values for the data points collected from 910 vertices in the LAGB showing in A and B. The red dashed line is a linear fit.

### 2.6.3 Segregation and precipitation at grain boundaries

Finally, a third case study was applied to our Auto-GB-CNN method. In this case, the APT dataset was taken from an Al-Zn-Mg-Cu alloy previously published by Zhao *et al.* [5]. Our aim now is to provide a statistical analysis revealing the solute distribution along the GB planes. Zhao *et al.* reported that magnesium (Mg) atoms segregate toward the GBs in this alloy during annealing and that this segregation contributed to the formation of precipitates along the GB plane. Using our ML method and mapping analysis, we can detect these solutes and provide a statistical analysis of them on these planes in an unbiased manner. Fig. 11 is the composition data measured along the GB plane with the area fraction of a given solute content given as a function of solute composition. As shown in this figure, for the as-quenched sample, two compositional peaks could be easily distinguished with the distribution of the Mg solute revealing it was not a Gaussian distribution, but rather a combination of multiple peaks as pointed out by the red arrows. These two peaks correspond to regions with different levels of solute partitioning. During the heat treatment, the peak with the lower solute content, approximate 2 at.% Mg, shifted to an even lower value, approximate 1 at.% Mg. This peak shift corresponded to the emergence of a solute-depleted region along the GB plane. The inset concentration maps in this figure are the GB planes that were identified by our Auto-GB-CNN method where one can visually note the solute evolution. We found that the peak at approximate 5 at.% Mg also shifted to lower solute content. However, with increasing heat treatment time, a longer compositional tail appears on the high solute content side. This longer tail corresponded to the center of the clusters that now are readily observed to form on the GB plane. These clusters act as a nucleus and assist the precipitation process along the GB plane as noted by Zhao *et al.* [5]. After increasing the annealing time to 24 h, the Mg content increased to a maximum of approximate 35 at.%. The area highlighted in the red dashed box in the compositional tail at 24 h shows the precipitates along the GB plane. Here, our Auto-GB-CNN method presented a reproducible and highly automated approach to the analysis to a time sequence study of APT experimental data. In doing so, new insights into interface thermodynamics and dynamics can be developed by having a reliable means of experimental analysis with no bias.

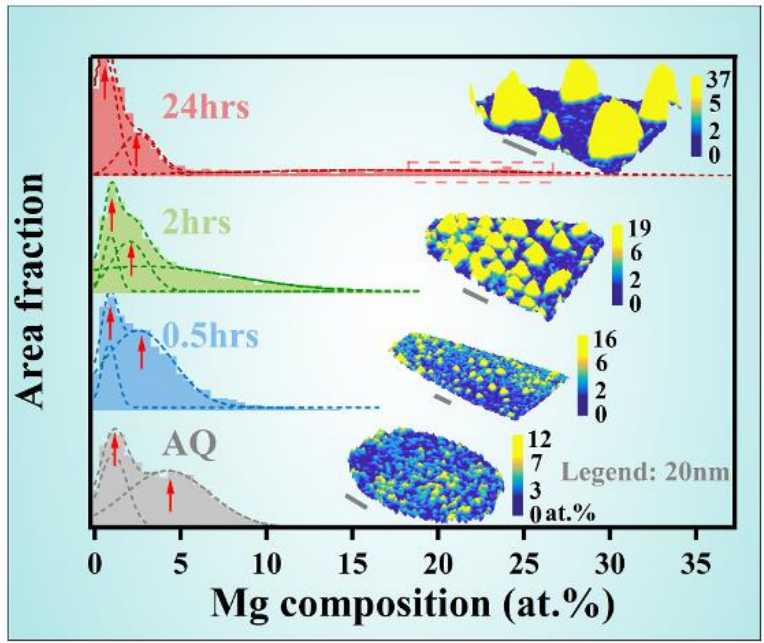

**Figure 11** The distribution of solute Mg atoms along the GB planes of a model Al-Zn-Mg-Cu alloy [5], represented by the area fraction of a given Mg content as a function of Mg composition. Dashed lines represent fittings from multiple peaks with Gaussian distribution. The embedded images are contour maps of solute composition along the GB plane of samples heat-treated at 120°C for different times.

## 3. Conclusions

This paper introduced a machine learning (ML)-based approach for mapping grain boundary (GB) composition and quantifying solute interfacial excess (IE) from atom probe tomography data. This methodology leads to an automatic way to reveal solute distribution along the GB plane, allowing the insights of distinguishable solute segregation behavior that extends beyond the dilute limit, where solute-solute interactions are non-negligible. The interaction between segregated species as well as segregated species with local defect structure can lead to phenomena, such as GB-specific segregation features as a function of misorientation and coincidence site lattice character, in-plane GB-spinodal decomposition, solute-dislocation interaction leading to complex segregation patterns, and the temporal evolution from early solute-GB segregation in the dilute limit to interfacial phase separation.

From the methodology aspect, we demonstrated our method using crystalline grains with near-columnar structures where the solute segregation is strongly coupled to the GB network. A CNN-based approach helped determine the local features of the grain interior, GB, and the triple junctions in profiling the GB network. Compositional mapping was achieved after meshing the GB network and performing subsequent quantification steps. With the help of the ladder diagram for each node in the GB mesh, automatic interfacial excess mapping was achieved as well as a means to determine the accurate location of a node for quantifying the local composition of a GB. We validated our implementation and approach with a synthesized data set. Thereafter, we applied this approach to three experimental case studies.

The advantage of the ML to APT dataset analysis is multifold. It can accelerate GB identification within the dataset by training the system to identify such features. As a result, it reduces the arduous, manual process of finding these features. Furthermore, it removes intrinsic and extrinsic bias users have in the search for these features. Our approach also eliminates the use of prior methods, such as iso-composition surfaces to identify the GBs, as this often depends on the solute being spatially and compositionally uniform over the GB plane, something that is rare. As a result, gaps exist on the actual plane. Using a ML approach, we identify the projected GB plane and corresponding GB network. This is then expanded to 3D and an automatic triangular mesh is applied that captures the boundary curvature and solute on such planes. With the GBs identified, we demonstrated in automatic means for quantifying GB segregation that increases the speed of data output, removes mutinous tasks, and, again, eliminates human bias in how the data is analyzed. Through this approach, atom probe information can become more readily reconstructed and analyzed that in helping to address a variety of physical and metallurgical factors that govern alloy design. With the success of is development in identifying 2D GB networks, future work aims to extend it to more complex polycrystals. The current approach is available as an open-source software package.

**Contribution statements**

X.Z. developed the codes for the mapping program. G.B.T. encouraged X.Z. to investigate the mapping methodology, provided two of the three experimental data sets to compare, and assisted in editing the manuscript draft. X.Z. and Y.W. conceived the ideal of the auto-detection algorithm. M.K. helped with tailoring, discussing, and optimizing the implementation of the algorithms. G.B.T., F.V., and R.D.K. gave suggestions on the mapping algorithm. F.V. and H.Z. provided the atom probe tomography data and assisted in data analysis. B.G. helped revise the manuscript. D.R. contributed to the discussion, construction and editing of the manuscript. All authors provided critical feedback and helped shape the research, algorithm, and manuscript.

**Acknowledgments**

X.Z. acknowledges the support from Alexander von Humboldt Foundation. X.Z. and G.B.T. gratefully recognize the National Science Foundation-DMR-1709803 for support of this research. We also thank Prof. Dr. Peter Felfer, Dr. Anna Ceguerra, and Ms. Varvara Efremova for opening their codes. X.Z thanks Dr. Alisson Kwiatkowski da Silva and Mr. Xinren Chen for the theoretical discussion on thermodynamics. We acknowledge Dr. Ting Luo for her assistance in testing our software package and for her suggestions to improve the usability.

**Data and code availability:** The current approach is available as an open-source software package (GitHub link: https://github.com/RhettZhou/APT_GB).

## References

[1] A.P. Sutton, R.W. Balluffi, Interfaces in Crystalline Materials, Clarendon Press, The University of Michigan, 1995.
[2] L. Lu, Y. Shen, X. Chen, L. Qian, K. Lu, Ultrahigh Strength and High Electrical Conductivity in Copper, Science 304 (2004) 422-426.
[3] D.A. Porter, K.E. Easterling, M. Sherif, Phase Transformations in Metals and Alloys, Third Edition (Revised Reprint), CRC Press, New York, 2009.
[4] F. Abdeljawad, D.L. Medlin, J.A. Zimmerman, K. Hattar, S.M. Foiles, A diffuse interface model of grain boundary faceting, J. Appl. Phys. 119(23) (2016) 235306.
[5] H. Zhao, F. De Geuser, A.K. da Silva, A. Szczepaniak, B. Gault, D. Ponge, D. Raabe, Segregation assisted grain boundary precipitation in a model Al-Zn-Mg-Cu alloy, Acta Mater 156 (2018) 318-329.
[6] L. Yao, S.P. Ringer, J.M. Cairney, M.K. Miller, The anatomy of grain boundaries: Their structure and atomic-level solute distribution, Scr. Mater. 69(8) (2013) 622-625.
[7] C.H. Liebscher, A. Stoffers, M. Alam, L. Lymperakis, O. Cojocaru-Mirédin, B. Gault, J. Neugebauer, G. Dehm, C. Scheu, D. Raabe, Strain-Induced Asymmetric Line Segregation at Faceted Si Grain Boundaries, Phys. Rev. Lett. 121(1) (2018) 015702.
[8] T. Meiners, T. Frolov, R.E. Rudd, G. Dehm, C.H. Liebscher, Observations of grain-boundary phase transformations in an elemental metal, Nature 579(7799) (2020) 375-378.
[9] E.A. Lazar, J.K. Mason, R.D. MacPherson, D.J. Srolovitz, Complete Topology of Cells, Grains, and Bubbles in Three-Dimensional Microstructures, Phys. Rev. Lett. 109(9) (2012) 095505.
[10] E.A. Lazar, J.K. Mason, R.D. MacPherson, D.J. Srolovitz, Distribution of Topological Types in Grain-Growth Microstructures, Phys. Rev. Lett. 125(1) (2020) 015501.
[11] E.A. Lazar, J. Han, D.J. Srolovitz, Topological framework for local structure analysis in condensed matter, P Natl Acad Sci USA 112(43) (2015) E5769-E5776.
[12] E.R. Homer, S. Patala, J.L. Priedeman, Grain Boundary Plane Orientation Fundamental Zones and Structure-Property Relationships, Sci. Rep. 5 (2015) 15476.
[13] X. Zhou, X.-x. Yu, T. Kaub, R.L. Martens, G.B. Thompson, Grain Boundary Specific Segregation in Nanocrystalline Fe(Cr), Sci. Rep. 6 (2016) 34642.
[14] M. Herbig, D. Raabe, Y.J. Li, P. Choi, S. Zaefferer, S. Goto, Atomic-Scale Quantification of Grain Boundary Segregation in Nanocrystalline Material, Phys. Rev. Lett. 112(12) (2014) 126103.
[15] J. Weissmüller, Alloy effects in nanostructures, Nanostruct. Mater. 3 (1993) 261-272.
[16] R. Kirchheim, Grain coarsening inhibited by solute segregation, Acta Mater 50(2) (2002) 413-419.
[17] T. Chookajorn, H.A. Murdoch, C.A. Schuh, Design of Stable Nanocrystalline Alloys, Science 337(6097) (2012) 951-954.
[18] J.W. Gibbs, The collected works of J.Willard Gibbs, Longmans, Green and Co., New York, 1902.

[19] R. Fowler, E.A. Guggenheim, Statistical Thermodynamics, Cambridge University Press, Cambridge, 1952.
[20] E.W. Hart, Two-Dimensional Phase Transformation in Grain Boundaries, Scr. Metall. 2(3) (1968) 179-182.
[21] M. Guttman, The role of residuals and alloying elements in temper embrittlement, 295(1413) (1980) 169-196.
[22] M. Yamaguchi, M. Shiga, H. Kaburaki, Grain boundary decohesion by impurity segregation in a nickel-sulfur system, Science 307(5708) (2005) 393-397.
[23] G. Svenningsen, M.H. Larsen, J.C. Walmsley, J.H. Nordlien, K. Nisancioglu, Effect of artificial aging on intergranular corrosion of extruded AlMgSi alloy with small Cu content, Corrosion Science 48(6) (2006) 1528-1543.
[24] R.Q. Wu, A.J. Freeman, G.B. Olson, First Principles Determination of the Effects of Phosphorus and Boron on Iron Grain-Boundary Cohesion, Science 265(5170) (1994) 376-380.
[25] J.W. Cahn, The Impurity-drag Effect in Grain Boundary Motion, Acta Mater 10 (1962) 789-798.
[26] K. Darling, M. Rajagopalan, M. Komarasamy, M. Bhatia, B. Hornbuckle, R. Mishra, K. Solanki, Extreme creep resistance in a microstructurally stable nanocrystalline alloy, Nature 537(7620) (2016) 378-381.
[27] H.A. Murdoch, C.A. Schuh, Stability of binary nanocrystalline alloys against grain growth and phase separation, Acta Mater 61(6) (2013) 2121-2132.
[28] H.A. Murdoch, C.A. Schuh, Estimation of grain boundary segregation enthalpy and its role in stable nanocrystalline alloy design, J. Mater. Res. 28(16) (2013) 2154-2163.
[29] D. Raabe, M. Herbig, S. Sandlöbes, Y. Li, D. Tytko, M. Kuzmina, D. Ponge, P.P. Choi, Grain boundary segregation engineering in metallic alloys: A pathway to the design of interfaces, Current Opinion in Solid State and Materials Science 18(4) (2014) 253-261.
[30] M.R. He, S.K. Samudrala, G. Kim, P.J. Felfer, A.J. Breen, J.M. Cairney, D.S. Gianola, Linking stress-driven microstructural evolution in nanocrystalline aluminium with grain boundary doping of oxygen, Nat. Commun. 7 (2016) 11225.
[31] A. Khalajhedayati, Z.L. Pan, T.J. Rupert, Manipulating the interfacial structure of nanomaterials to achieve a unique combination of strength and ductility, Nat. Commun. 7 (2016) 10802.
[32] M. Jahazi, J.J. Jonas, The non-equilibrium segregation of boron on original and moving austenite grain boundaries, Mater. Sci. Eng., A 335(1) (2002) 49-61.
[33] S.A. Kube, W.Y. Xing, A. Kalidindi, S. Sohn, A. Datye, D. Amram, C.A. Schuh, J. Schroers, Combinatorial study of thermal stability in ternary nanocrystalline alloys, Acta Mater 188 (2020) 40-48.
[34] D. Raabe, S. Sandlöbes, J. Millán, D. Ponge, H. Assadi, M. Herbig, P.P. Choi, Segregation engineering enables nanoscale martensite to austenite phase transformation at grain boundaries: A pathway to ductile martensite, Acta Mater 61(16) (2013) 6132-6152.
[35] P. Lejcek, S. Hofmann, Thermodynamics and Structural Aspects of Grain-Boundary Segregation, Critical Reviews in Solid State and Materials Sciences 20(1) (1995) 1-85.
[36] P. Lejcek, S. Hofmann, V. Paidar, Solute segregation and classification of [100] tilt grain boundaries in alpha-iron: consequences for grain boundary engineering, Acta Mater 51(13) (2003) 3951-3963.
[37] P. Lejcek, S. Hofmann, Thermodynamics of grain boundary segregation and applications to anisotropy, compensation effect and prediction, Critical Reviews in Solid State and Materials Sciences 33(2) (2008) 133-163.
[38] A. Kwiatkowski da Silva, R.D. Kamachali, D. Ponge, B. Gault, J. Neugebauer, D. Raabe, Thermodynamics of grain boundary segregation, interfacial spinodal and their relevance for nucleation during solid-solid phase transitions, Acta Mater 168 (2019) 109-120.

[39] L.L. Li, Z.M. Li, A.K. da Silva, Z.R. Peng, H. Zhao, B. Gault, D. Raabe, Segregation-driven grain boundary spinodal decomposition as a pathway for phase nucleation in a high-entropy alloy, Acta Mater 178 (2019) 1-9.

[40] A.K. da Silva, D. Ponge, Z. Peng, G. Inden, Y. Lu, A. Breen, B. Gault, D. Raabe, Phase nucleation through confined spinodal fluctuations at crystal defects evidenced in Fe-Mn alloys, Nat. Commun. 9 (2018) 1137.

[41] A.K. da Silva, R.D. Kamachali, D. Ponge, B. Gault, J. Neugebauer, D. Raabe, Thermodynamics of grain boundary segregation, interfacial spinodal and their relevance for nucleation during solid-solid phase transitions, Acta Mater 168 (2019) 109-120.

[42] P. Wynblatt, D. Chatain, Solid-state wetting transitions at grain boundaries, Mater. Sci. Eng., A 495(1-2) (2008) 119-125.

[43] P. Lejcek, Grain Boundary Segregation in Metals, 2010.

[44] D. Udler, D.N. Seidman, Solute-Atom Segregation at (002) Twist Boundaries in Dilute Ni-Pt Alloys - Structural Chemical Relations, Acta Metallurgica Et Materialia 42(6) (1994) 1959-1972.

[45] M. Wagih, C.A. Schuh, Grain boundary segregation beyond the dilute limit: Separating the two contributions of site spectrality and solute interactions, Acta Mater 199 (2020) 63-72.

[46] C. Scheu, Electron energy-loss near-edge structure studies at the atomic level: reliability of the spatial difference technique, J Microsc-Oxford 207 (2002) 52-57.

[47] Y. Ikuhara, P. Thavorniti, T. Sakuma, Solute segregation at grain boundaries in superplastic $SiO_2$-doped TZP, Acta Mater 45(12) (1997) 5275-5284.

[48] C. O'Brien, C. Barr, P. Price, K. Hattar, S. Foiles, Grain boundary phase transformations in PtAu and relevance to thermal stabilization of bulk nanocrystalline metals, J. Mater. Sci. 53(4) (2018) 2911-2927.

[49] B. Feng, T. Yokoi, A. Kumamoto, M. Yoshiya, Y. Ikuhara, N. Shibata, Atomically ordered solute segregation behaviour in an oxide grain boundary, Nat. Commun. 7(1) (2016) 11079.

[50] D.J. Larson, B. Gault, B.P. Geiser, F. De Geuser, F. Vurpillot, Atom probe tomography spatial reconstruction: Status and directions, Current Opinion in Solid State & Materials Science 17(5) (2013) 236-247.

[51] D. Amram, C.A. Schuh, Interplay between thermodynamic and kinetic stabilization mechanisms in nanocrystalline Fe-Mg alloys, Acta Mater 144 (2018) 447-458.

[52] S.-I. Baik, M.J. Olszta, S.M. Bruemmer, D.N. Seidman, Grain-boundary structure and segregation behavior in a nickel-base stainless alloy, Scr. Mater. 66(10) (2012) 809-812.

[53] S.K. Samudrala, P.J. Felfer, V.J. Araullo-Peters, Y. Cao, X.Z. Liao, J.M. Cairney, New atom probe approaches to studying segregation in nanocrystalline materials, Ultramicroscopy 132 (2013) 158-63.

[54] Z.R. Peng, Y.F. Lu, C. Hatzoglou, A.K. da Silva, F. Vurpillot, D. Ponge, D. Raabe, B. Gault, An Automated Computational Approach for Complete In-Plane Compositional Interface Analysis by Atom Probe Tomography, Microsc. Microanal. 25(2) (2019) 389-400.

[55] P. Felfer, B. Scherrer, J. Demeulemeester, W. Vandervorst, J.M. Cairney, Mapping interfacial excess in atom probe data, Ultramicroscopy 159 (2015) 438-444.

[56] D.N. Seidman, B.W. Krakauer, D. Udler, Atomic scale studies of solute-atom segregation at grain boundaries: Experiments and simulations, J. Phys. Chem. Solids 55(10) (1994) 1035-1057.

[57] R. Xu, C.C. Chen, L. Wu, M.C. Scott, W. Theis, C. Ophus, M. Bartels, Y. Yang, H. Ramezani-Dakhel, M.R. Sawaya, H. Heinz, L.D. Marks, P. Ercius, J.W. Miao, Three-dimensional coordinates of individual atoms in materials revealed by electron tomography, Nat. Mater. 14(11) (2015) 1099-1103.

[58] O.C. Hellman, J.B.d. Rivage, D.N. Seidman, Efficient sampling for three-dimensional atom probe microscopy data, Ultramicroscopy 95 (2003) 199-205.

[59] W.E. Lorensen, H.E. Cline, Marching cubes: A high resolution 3D surface construction algorithm, Proceedings of the 14th annual conference on Computer graphics and interactive techniques, Association for Computing Machinery, 1987, pp. 163–169.

[60] B.M. Jenkins, F. Danoix, M. Gouné, P.A.J. Bagot, Z. Peng, M.P. Moody, B. Gault, Reflections on the Analysis of Interfaces and Grain Boundaries by Atom Probe Tomography, Microsc. Microanal. 26(2) (2020) 247-257.
[61] B. Gault, M.P. Moody, J.M. Cairney, S.P. Ringer, Atom Probe Microscopy, Springer New York, New York, 2012.
[62] O.C. Hellman, J.A. Vandenbroucke, J. Rüsing, D. Isheim, D.N. Seidman, Analysis of Three-dimensional Atom-probe Data by the Proximity Histogram, Microsc. Microanal. 6(5) (2000) 437-444.
[63] L. Yao, M.P. Moody, J.M. Cairney, D. Haley, A.V. Ceguerra, C. Zhu, S.P. Ringer, Crystallographic structural analysis in atom probe microscopy via 3D Hough transformation, Ultramicroscopy 111(6) (2011) 458-63.
[64] P. Felfer, A. Ceguerra, S. Ringer, J. Cairney, Applying computational geometry techniques for advanced feature analysis in atom probe data, Ultramicroscopy 132 (2013) 100-106.
[65] P. Felfer, P. Benndorf, A. Masters, T. Maschmeyer, J.M. Cairney, Revealing the Distribution of the Atoms within Individual Bimetallic Catalyst Nanoparticles, Angew. Chem. Int. Edit. 53(42) (2014) 11190-11193.
[66] B.W. Krakauer, D.N. Seidman, Absolute atomic-scale measurements of the Gibbsian interfacial excess of solute at internal interfaces, Phys. Rev. B 48(9) (1993) 6724-6727.
[67] Y. Freund, R.E. Schapire, A Decision-Theoretic Generalization of On-Line Learning and an Application to Boosting, Journal of Computer and System Sciences 55(1) (1997) 119-139.
[68] Y. Wei, Z. Peng, M. Kühbach, A. Breen, M. Legros, M. Larranaga, F. Mompiou, B. Gault, 3D nanostructural characterisation of grain boundaries in atom probe data utilising machine learning methods, PLOS ONE 14(11) (2019) e0225041.
[69] M. Kühbach, M. Kasemer, B. Gault, A. Breen, On Open and Strong-Scaling Tools for Atom Probe Crystallography: High-Throughput Methods for Indexing Crystal Structure and Orientation, Preprint arXiv: 2009.00735 (2020).
[70] L. Li, R. Darvishi Kamachali, Z. Li, Z. Zhang, Grain boundary energy effect on grain boundary segregation in an equiatomic high-entropy alloy, Phys. Rev. Mater. 4(5) (2020) 053603.
[71] T.L. Martin, A. Radecka, L. Sun, T. Simm, D. Dye, K. Perkins, B. Gault, M.P. Moody, P.A.J. Bagot, Insights into microstructural interfaces in aerospace alloys characterised by atom probe tomography, Mater Sci Tech-Lond 32(3) (2016) 232-241.
[72] Y. LeCun, Y. Bengio, G. Hinton, Deep learning, Nature 521(7553) (2015) 436-444.
[73] R.O. Duda, P.E. Hart, Use of the Hough transformation to detect lines and curves in pictures, Communications of the ACM 15(Commun. ACM) (1972) 11–15.
[74] W. Zhang, K. Itoh, J. Tanida, Y. Ichioka, Parallel distributed processing model with local space-invariant interconnections and its optical architecture, Appl. Opt. 29(32) (1990) 4790-4797.
[75] M. Rastegari, V. Ordonez, J. Redmon, A. Farhadi, XNOR-Net: ImageNet Classification Using Binary Convolutional Neural Networks, European conference on computer vision, Springer International Publishing, Cham, 2016, pp. 525-542.
[76] Y. Wei, B. Gault, R.S. Varanasi, D. Raabe, M. Herbig, A.J. Breen, Machine-learning-based atom probe crystallographic analysis, Ultramicroscopy 194 (2018) 15-24.
[77] S. Madireddy, D.W. Chung, T. Loeffler, S.K.R.S. Sankaranarayanan, D.N. Seidman, P. Balaprakash, O. Heinonen, Phase Segmentation in Atom-Probe Tomography Using Deep Learning-Based Edge Detection, Sci. Rep. 9 (2019) 20140.
[78] J. Zelenty, A. Dahl, J. Hyde, G.D.W. Smith, M.P. Moody, Detecting Clusters in Atom Probe Data with Gaussian Mixture Models, Microsc. Microanal. 23(2) (2017) 269-278.
[79] I. Mouton, S. Katnagallu, S.K. Makineni, O. Cojocaru-Mirédin, T. Schwarz, L.T. Stephenson, D. Raabe, B. Gault, Calibration of Atom Probe Tomography Reconstructions Through Correlation with Electron Micrographs, Microsc. Microanal. 25(2) (2019) 301-308.

[80] A.C. Day, H. Francois-Saint-Cyr, B.P. Geiser, T. Payne, E. Oltman, S.P. Ringer, D.A. Reinhard, Recent Developments in APT Analysis Automation and Support for User-Defined Custom Analysis Procedures in IVAS 4, Microsc. Microanal. 25(S2) (2019) 338-339.
[81] B.O. Community, Blender - a 3D modeling and rendering package, Amsterdam, 2018.
[82] D.J. Larson, T.J. Prosa, R.M. Ulfig, B.P. Geiser, T.F. Kelly, Local Electrode Atom Probe Tomography: A User's Guide, Springer, New York, 2013.
[83] X. Zhou, X. Yu, D. Jacobson, G.B. Thompson, A molecular dynamics study on stress generation during thin film growth, Appl. Surf. Sci. 469 (2019) 537-552.
[84] M. Miller, K., The effects on local magnification and trajectory aberrations on atom probe analysis, J. Phys. Colloques 48(C6) (1987) C6-565-570.
[85] Y. Lecun, L.D. Jackel, B. Boser, J.S. Denker, H.P. Graf, I. Guyon, D. Henderson, R.E. Howard, W. Hubbard, Handwritten Digit Recognition - Applications of Neural Network Chips and Automatic Learning, Ieee Commun Mag 27(11) (1989) 41-46.
[86] Y. Lecun, L. Bottou, Y. Bengio, P. Haffner, Gradient-based learning applied to document recognition, P Ieee 86(11) (1998) 2278-2324.
[87] A. Krizhevsky, I. Sutskever, G.E. Hinton, Imagenet classification with deep convolutional neural networks, Advances in neural information processing systems, 2012, pp. 1097-1105.
[88] B. Gault, M.P. Moody, F. De Geuser, A. La Fontaine, L.T. Stephenson, D. Haley, S.P. Ringer, Spatial resolution in atom probe tomography, Microsc. Microanal. 16(1) (2010) 99-110.
[89] M. Miller, M. Hetherington, Local magnification effects in the atom probe, Surf. Sci. 246(1-3) (1991) 442-449.
[90] M. Abadi, A. Agarwal, P. Barham, E. Brevdo, Z. Chen, C. Citro, G.S. Corrado, A. Davis, J. Dean, M. Devin, S. Ghemawat, I. Goodfellow, A. Harp, G. Irving, M. Isard, R. Jozefowicz, Y. Jia, L. Kaiser, M. Kudlur, J. Levenberg, D. Mané, M. Schuster, R. Monga, S. Moore, D. Murray, C. Olah, J. Shlens, B. Steiner, I. Sutskever, K. Talwar, P. Tucker, V. Vanhoucke, V. Vasudevan, F. Viégas, O. Vinyals, P. Warden, M. Wattenberg, M. Wicke, Y. Yu, X. Zheng, TensorFlow: Large-scale machine learning on heterogeneous systems,  (2015).
[91] A. Gupta, X.Y. Zhou, G.B. Thompson, G.J. Tucker, Role of grain boundary character and its evolution on interfacial solute segregation behavior in nanocrystalline Ni-P, Acta Mater 190 (2020) 113-123.
[92] T.C. Lee, R.L. Kashyap, C.N. Chu, Building Skeleton Models via 3-D Medial Surface Axis Thinning Algorithms, CVGIP: Graphical Models and Image Processing 56(6) (1994) 462-478.
[93] H. Edelsbrunner, E.P. Mücke, Three-dimensional alpha shapes, ACM Transactions on Graphics 13(1) (1994) 43–72.
[94] S. Hagége, C.B. Carter, F. Cosandey, S.L. Sass, The variation of grain boundary structural width with misorientation angle and boundary plane, Philosophical Magazine A 45(4) (1982) 723-740.
[95] G. Gottstein, L.S. Shvindlerman, Grain Boundary Migration in Metals: Thermodynamics, Kinetics, Applications, Taylor & Francis, New York, 1999.
[96] R.E. Mistler, R.L. Coble, Grain‐boundary diffusion and boundary widths in metals and ceramics, J. Appl. Phys. 45(4) (1974) 1507-1509.
[97] M.R. Chellali, Z. Balogh, H. Bouchikhaoui, R. Schlesiger, P. Stender, L. Zheng, G. Schmitz, Triple Junction Transport and the Impact of Grain Boundary Width in Nanocrystalline Cu, Nano Letters 12(7) (2012) 3448-3454.
[98] R. Killick, P. Fearnhead, I.A. Eckley, Optimal Detection of Changepoints With a Linear Computational Cost, Journal of the American Statistical Association 107(500) (2012) 1590-1598.
[99] X. Zhou, A. Gupta, G.J. Tucker, G.B. Thompson, Manipulation of Solute Partitioning Mechanisms for Nanocrystalline Stability, Acta Mater 208 (2021) 116662.
[100] D. Hull, D.J. Bacon, Introduction to Dislocations (Fourth Edition), Butterworth-Heinemann, Oxford, 2001.

[101] X. Zhou, J. Mianroodi, A.K. da Silva, T. Koenig, G. Thompson, P. Shanthraj, D. Ponge, B. Gault, B. Svendsen, D. Raabe, The hidden structure dependence of the chemical life of dislocations, Sci Adv 7(16) (2021) eabf0563.
[102] V.J. Araullo-Peters, B. Gault, S.L. Shrestha, L. Yao, M.P. Moody, S.P. Ringer, J.M. Cairney, Atom probe crystallography: Atomic-scale 3-D orientation mapping, Scr. Mater. 66(11) (2012) 907-910.
[103] Y. Yu, S.Y. Zhang, A.M. Mio, B. Gault, A. Sheskin, C. Scheu, D. Raabe, F.Q. Zu, M. Wuttig, Y. Amouyal, O. Cojocaru-Miredin, Ag-Segregation to Dislocations in PbTe-Based Thermoelectric Materials, Acs Appl Mater Inter 10(4) (2018) 3609-3615.